\pdfoutput=1
\documentclass[a4paper]{article}
\title{A Method to Produce Intense Positron Beams via Electro Pair Production on Electrons }
\author{Berthold Schoch}
\date{July 13  2016}
\usepackage{graphicx}
\begin{document}
\maketitle
Physikalisches Institut, Universit\"at Bonn, Nussallee 12, D-53115 Bonn,
Germany 
\begin{abstract}
 Intense positron beams can be prepared via 
electro production with the reaction  $e^{-} + e^{-} \rightarrow e^{-}+e^{+}+e^{-}+e^{-}$ due to
the availability of high current electron beams. Head on
collisions inside of a magnetic field of a solenoid are used to produce unpolarized/polarized positrons via 
$e^{-}-e^{+}$ -  pair production. A flux of unpolarized positrons, $N_{positron}=6.55 \cdot 10^{11} s^{-1}$, has been obtained 
as a result of calculations for a special case
of parameters with beam currents for both electron beams of $I_{e}$=10 mA and beam momenta of $p_{beam 1}$=0.63 MeV/c
and $p_{beam 2}$=10.0 MeV/c. Intensities are an order of magnitude smaller in the case of polarized positrons  due to
a reduced current of one of the electron beams. Suitable cuts on momenta of the positrons allow to achieve a polarization
transfer from electrons to positrons of $\geq $ 85\% reducing, however, the intensity to 27\%.   
\end{abstract}
\section{Introduction}
Positrons and beams of positrons find applications in several branches of science.
Positrons as a probe are used in solid state
physics as well as in particle physics, covering an energy range from eV up
to 250 GeV, respectively.
The status of the field at low and medium energies and the need for improved positron
beams have been stressed in several recent publications \cite{voutier},\cite{golge},\cite{grames}. 
Ideally, the beam qualities should approach that of
electron beams and in certain cases, polarized beams are needed, too. So far, the
decay of radio isotopes via $\beta $-decay and electron-positron pair
production via photo production in the electric field of heavy nuclei 
have been the reactions of
choice. Positrons leaving beam pipes on nuclear power reactors provide, so
far, the most intense beams of thermalized positrons, however, in the process
of thermalization only a small fraction, of the order of $10^{-4}$, survive
from around $10^{13}$/s to $10^{14}$/s down to $10^{9}$/s to $10^{10}$/s.
Pair production on the other hand allows preparation of beams of
positrons over a wide range of energy. 
The accelerator facility in Saclay, France \cite{argan}, 
serves as an early example. Beams of positrons have been 
prepared in the 1980ties in Saclay, with a 720 MeV electron LINAC, for particle
physics experiments in the several hundred MeV region. 
Currents of 20-50 nA for a beam with a momentum spread of $\Delta $p/p=1.5\%
and an emittance of 4 mmmrad have been prepared via bremsstrahlung 
processes on high Z targets. An efficiency of 0.1 $\%$ has been reached 
to transform an incoming electron beam into a positron
beam. A similar set-up was used with the Mainz LINAC \cite{leicht} for positron
annihilation experiments. Both facilities were able 
to run with electron beam currents of the order of $I_{e}=50\mu A$.
Problems to overcome were the construction of high Z targets
standing the heat and build a shielding against a high flux of neutrons 
around those targets.
\begin{figure}[h!]
\centering
\includegraphics[scale=0.4]{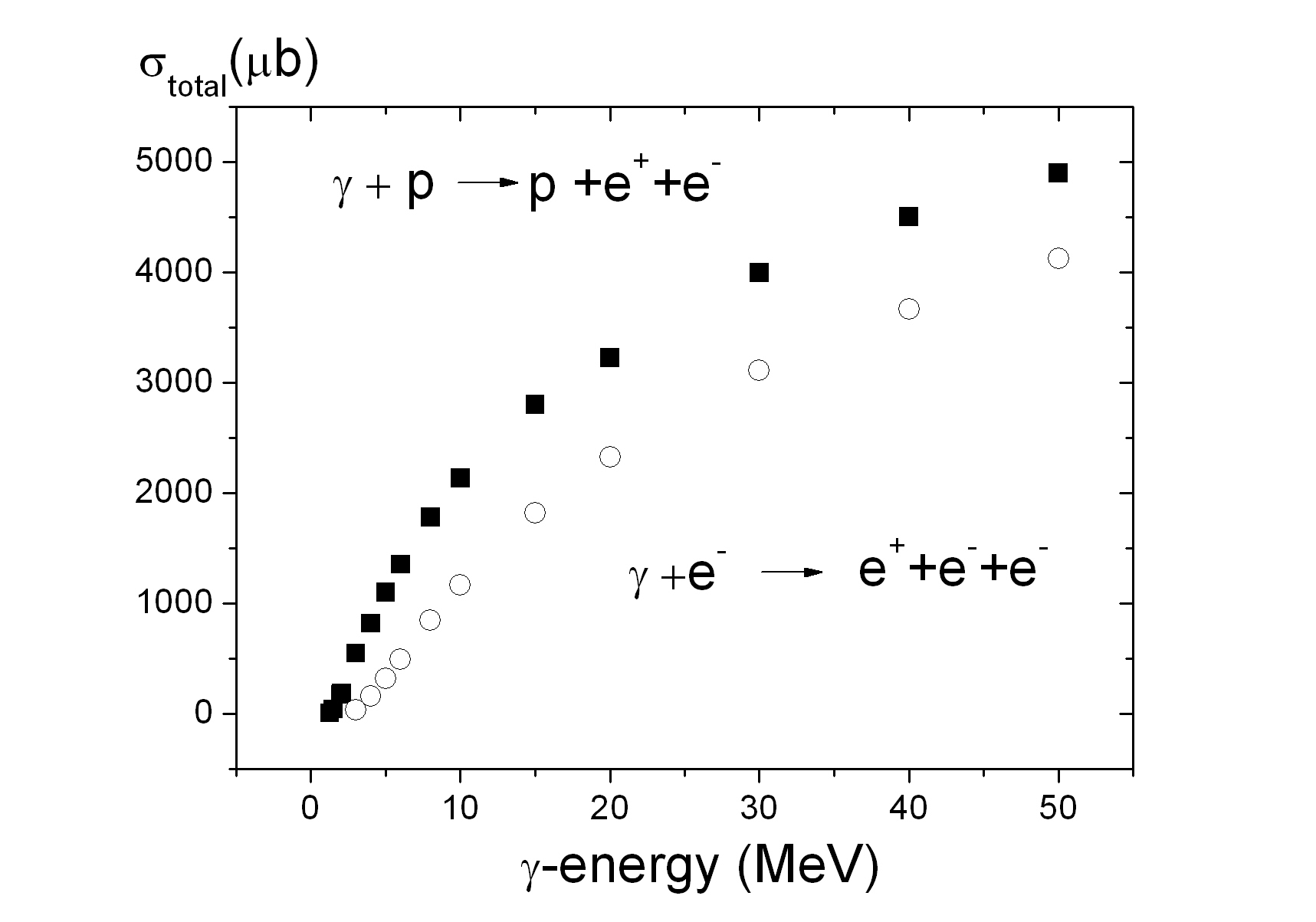}
\caption{$\gamma$-pair production cross sections in the Coulomb fields of proton and electron, data taken from \protect Ref. \cite{gimm}.}
\label{fig:2}
\end{figure}
\section{Electro production within a magnetic field of a solenoid}
\label{sec:2}
With the development of injectors providing polarized electron beams, 
pair production offers a convenient way to produce
polarized positrons via the use of circular polarized 
$\gamma $-rays.
 Improvements and a better exploitation of the method for future applications
are necessary and possible. In this paper the reaction of trident (triplet) pair
production is proposed as a source of a positron beam. Fig. 1 shows the
total cross section for positron production via the reaction $\gamma
+e^{-}\rightarrow e^{-}+ e^{+}+ e^{-}$ together with the total cross section of the reaction
$\gamma
+p\rightarrow e^{-}+ e^{+}+ p$
taken from Ref. \cite{gimm}. The similar sizes of the cross sections 
demonstrate that the Coulomb field
plays the decisive role for the reaction despite the large mass difference of the targets. The cross
section for photo production of $e^{+}-e^{-}$ pairs on electrons $\sigma _{total}^{pair}$=1 mb
at $E_{\gamma }$=10 MeV (see Fig. 1) has to be compared with the cross section
on high Z elements e.g. on Ta(Z=73) with $\sigma _{total}^{pair}$=10 b.
The difference explains the preferred use of high Z targets for
photo production of positrons. 
\vspace{1pt}
By using electrons as a target
special properties of the reaction and kinematic must 
be found and explored in order to compensate for the smallness of that cross
section. Instead of photo production electro production might offer a way 
to achieve a high enough luminosity for being competitive with other
methods. 
A schematic set-up for an electro production experiment is shown in Fig. 2.
Two electron beams collide head on and produce electron-positron pairs in a
longitudinal magnetic field within a solenoid.
\begin{figure}[h!]
\centering
\includegraphics[scale=0.5]{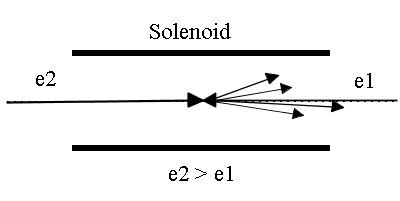}
\caption{Schematic picture of the set-up.}
\label{fig:2}
\end{figure}
\begin{figure}[h!]
\centering
\includegraphics[scale=0.4]{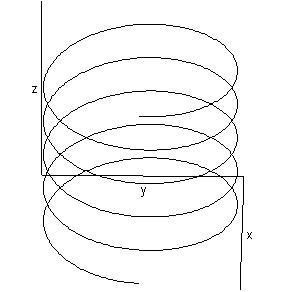}
\caption{The helical trajectory of an electron in the magnetic field of a solenoid after 
entering the field parallel to
the symmetry axis z at a distance R.}
\label{fig:3}
\end{figure}
\subsection{Luminosity}
\label{sec:3}
Luminosity L for such a set-up for one interaction zone is given by: 
\begin{equation}
L=\frac{N_{e2}\cdot N_{e1}}{4\pi \cdot \sigma _{x}\cdot \sigma _{y}}%
(cm^{-2}\cdot s^{-1})
\end{equation}
$N_{e1}$and $N_{e2}$ stand for the number of electrons/s in
the beams, respectively, and $\sigma _{x}$ and $\sigma _{y}$ for the
Gaussian beam profiles of the two intersecting beams of same extension.
D.C. beams of electrons are assumed in the following discussion but other  time
structures are not excluded. Electron currents in the mA range with useful emittances became
available in the last years e.g. by the development of energy recovering linacs.
The results of those developments
gave reason to consider a set-up as shown in Fig. 2 for a positron source with
remarkable features. However, cross
sections for electro production are generally two orders of
magnitude smaller compared to photo production cross sections investigating the same reactions
at the same energy transfers. Thus, the
beam extensions $\sigma _{x}$ and $\sigma _{y}$ at interaction zone
will be decisive whether a positron current of the order  of 10$^{10}$
positrons/s or more can be reached.
\subsection{Particle trajectories in a field of a solenoid  }
\label{sec:4}
Solenoid fields are used to focus charged particle beams in beam lines for many applications. The 
propagation of beams in beam lines with solenoids is well incorporated in beam transport programs. 
The trajectories of single
particles, however, are of special importance for the application of  extended solenoid fields providing
several foci and, thus, new opportunities for colliding beams see e.g. \cite{kumar}.
The trajectory of a charged particle entering the solenoid field parallel to the symmetry axis, z-direction,
with distance R to the symmetry axis is considered under the assumption of a fringe field with no extension 
(hard edge approximation) and incoming momentum large enough in order to leave the magnetic field with little 
change in its transverse coordinates.
Inside, the particle's trajectory forms a helix, see Fig. 3, propagating in z- direction  with radius R/2 
in the projected x-y plane whose center is at distance R/2 from the symmetry axis of the solenoid. The particle rotates with the cyclotron frequency 
$\omega $= $\frac{e\cdot B}{\gamma \cdot m}$ (rad/s$\cdot$T),  with e and m the charge and rest mass of the particle, $\gamma =\frac{E}{m}$ with E as total energy of the particle and B the magnetic field inside the solenoid. All 
particles rotate with the same frequency  independently of R.
\begin{figure}[h!]
\centering
\includegraphics[scale=0.5]{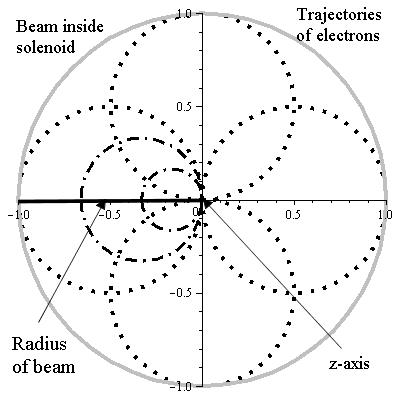}
\caption{x-y projection of the trajectories of 4 particles which started at z=$\Delta z \rightarrow 0$ at distance R from
the axis and two particles at a distance 2/3 R and 1/3 R, respectively, after one period of cyclotron frequency.}
\label{fig:4}
\end{figure}
\begin{figure}[h!]
\centering
\includegraphics[scale=0.7]{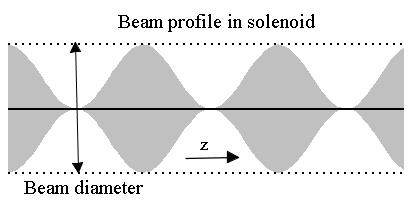}
\caption{Beam modulation in B-field along the z-axis of the solenoid for particles on the periphery, distance from
z-axis, of the incoming beam.}
\label{fig:5}
\end{figure}
Fig. 4 illustrates the projection of the rotation of six particles onto the x-y plane in one period, demonstrating the focusing
to the symmetry axis of the solenoid. That means that for an exactly parallel beam of particles with diameter 2R outside the field and a constant density along the z-axis a density modulated beam traverses the solenoid as shown in Fig. 5. for the outer four particles.
The inner two particles reach at the same time the axis as the outer ones but have slightly different velocities in z-direction because the radii of the helices are different. A discussion of all relevant quantities for fixed beam parameters can
be found in the appendix.
The conservation of canonical angular momentum 
\begin{equation}
\overrightarrow{L}=\overrightarrow{r}\times (\overrightarrow{P}+e\cdot 
\overrightarrow{A})
\end{equation}
serves as a starting point for discussion of trajectories of electrons, with \overrightarrow{A} as vector potential and \overrightarrow{P}
as the momentum of electrons. L=0 stands for a beam outside of the magnetic field parallel to the symmetry axis. Due to induced Lenz force that value remains zero inside. Quantitatively, a solenoid magnet with field $B_{z}$ has a vector potential $A_{\phi}$=$\frac{r\cdot B_{z}}{2}$, with $\phi$ being the angle of rotation in x-y plane. A current (due to Lenz law)
is induced in such a way that r$\cdot (P_{\phi}+e \cdot A_{\phi})=0$. A real beam has, however, transverse momenta due to
finite emittances and, thus changing somewhat
the ideal picture. 
One of the challenges for the preparation of a beam  will be to optimize the parameters in view
to achieve high luminosity.  
Another important relation carrying out calculations is given by the radius of an electron in a 
homogeneous magnetic field 
$r(m)=\frac{p(MeV/c)}{300\cdot B(Tesla)}$.
With special choices of beam and field parameters a case study will be presented
in order to prepare a discussion of the luminosity.
\begin{figure}[h!]
\centering
\includegraphics[scale=0.7]{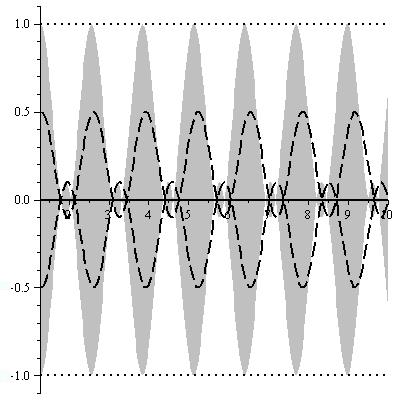}
\caption{Beam modulation in B-field of  the solenoid for particle positions on the periphery and half way of the radius of the incoming beam. Both trajectories include the axis of symmetry by $\Delta r=100 \mu $m in order to amplify  the effect.}
\label{fig:6}
\end{figure}
\subsection{A case study }
\label{sec:5}
Two electron beams are brought to interaction in a 
set-up as shown schematically in Fig. 2. Electron beam e2
with momentum $p_{2}$=10 MeV/c interacts in a 
longitudinal magnetic field with a field strength of 1.0 Tesla with 
electron beam e1 with momentum 0.63 MeV/c  corresponding to a kinetic energy
of 300 keV. 
In order to reach relevant numbers for luminosity detailed calculations of the trajectories 
of the beams in the solenoid have to be performed.
The results are presented in the appendix.
As so far discussed, all particle trajectories pass through the magnetic axis of the solenoid. However, the velocities
in z-direction of 
particle trajectories change due to their r-dependence. 
The size of the effect
finds its expression on the different path lengths s of the helices, e.g. for peripheral particles, with s=1.032 m
and s=1.0001 m for the e1- and e2 beam, respectively. 
The number of foci are for the e1-beam 78 and 5 for the e2-beam. 
Remains to be discussed the waists of the beams finding its  expression in the formula 
for the luminosity by $\sigma _{x}\cdot \sigma _{y}$ located at the foci which
are shown in Fig. 5 for a trajectory of electrons entering the magnetic field coming from
the periphery of the beam. It stands for all electrons entering the magnetic field parallel
to the axis of the field. However, such an ideal situation can only be approached by optimizing
the available parameters, like the incoming momenta and the emittances of the particles involved, 
the strength of the magnetic field and the diameter of the beam.
Unavoidable transverse components of the incoming beams lead to trajectories including or excluding the axis
of field symmetry and, thus, widening the waists of beams. Non zero values of emittances lead to transverse momentum 
components of the beams. Focusing and defocusing of beams lead to transverse momentum components, too.
Fig. 6 shows the beam profiles for particles situated on the periphery of the beam with r=1 mm and half way to the periphery with r= 0.5 mm.
A inclusion of the axis of symmetry by a distance of $\Delta r=100 \mu m$ has been assumed in order to amplify 
the effect of a transverse component of momentum responsible for waists of the beams. 
In the case study outlined in the appendix values for the two beams 
of $\Delta r=13 \mu m$ and $\Delta r=17 \mu m$ have been calculated for beam e1 and beam e2, respectively. However, as input for 
formula 1 defining the luminosity a value for $\sigma _{x}$ as well as $\sigma _{y}$ of $\Delta r=30 \mu m$ will be taken
and, thus, including possible uncertainties in calibrating the set-up. Experience from other set-ups give confidence
that calibrations with a precision of the order $5 \mu m$ can be achieved. 
Fig. 6 shows in addition that the velocities in z-direction are r-dependent, as mentioned above. However,
that behavior of the beam particles has no relevance for the application considered in this paper.
The number of foci is another important outcome of the calculations presented in the Appendix. 
Five foci of beam e2 and 78 foci of bam e1 demonstrate the usefulness of a solenoid to create many interaction regions
for two beams in head on collisions.
\subsection{Positron production with electrons }
\label{sec:6}
\subsubsection{Thresholds}
\label{sec:7}
The threshold energy $E_{e2}$  for the reaction $e_{e2}^{-}+e_{e1}^{-}%
\rightarrow e^{-}+e^{+}+e^{-}$ $+e^{-}$ with e1 at rest in the lab. system yields
$E_{e2}$ $=7\cdot m_{e}$ according to the equation for invariant masses
\begin{equation}
m_{invariant}^{2}=(4\cdot m_{e})^{2}=(E_{e2}+E_{e1})^{2}-(\overrightarrow{p}%
_{e2}-\overrightarrow{p}_{e1})^{2}
\end{equation}
The threshold energy $E_{e2}$  for a beam $e2$ with a head on collision with
beam $e1$ with energy $E_{e2}=0.811 MeV$
and momentum $p_{e1}=0.63 MeV/c$ results to $E_{e2}=1.36 MeV$ or $\allowbreak
2.\,\allowbreak 66\cdot m_{e}.$
Those energies should be compared with the c.m. energy \ $%
E_{cm}^{e2+e1}=5.41 MeV$ for the chosen 
e2 and e1 energies.
\subsubsection{Kinematic of positron momenta}
\label{sec:8}
The reaction considered has two particles in the initial state and four particles in the final state all with 
the same mass. The available momentum space for positrons can be determined by reducing the four
body state to a two body state considering the c.m. movements of two pairs of particles consisting of 
$e^{-}+e^{+}$ and $e^{-}+e^{-}$. From those distributions it is possible to extract the available momentum
space of the positrons. Fig. 7 shows that momentum space. 
The positrons are almost all emitted into forward direction with the highest momentum $p_{max}$ reaching almost $p_{max}$=5 MeV/c.
The transverse momentum with $p_{max}^{transverse}$=1.25 MeV/c leads to a helical trajectory with a radius r of $r_{max}$=4.2 mm. 
\begin{figure}[h!]
\centering
\includegraphics[scale=0.6]{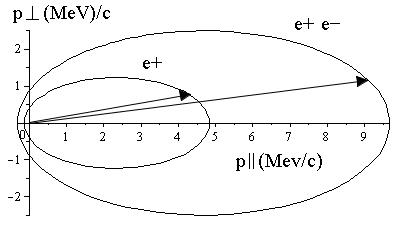}
\caption{The momentum space of positrons as well as the pairs of $e^{-}+e^{+}$ as  $e^{-}+e^{-}$.}
\label{fig:7}
\end{figure}
\subsubsection{Energy of beam e2 in the rest system of beam e1}
\label{sec:9}
The cross sections are usually measured in the laboratory system with e1 at rest.That is the case shown in Fig. 1 for $ \gamma $ - pair production cross sections. In head on collisions, however,
smaller energies are needed to cover the same energy region in the c.m.-system. The lab. energy $E_{e2}^{e1(rest)}$ is given by a 
Lorentz transformation 
\begin{equation}
E_{e2}^{e1(rest)}= \gamma _{e1} \cdot (E_{e2}+ \beta _{e1} \cdot p_{e2}) 
\end{equation}
yields $E_{e2}^{e1(rest)}$=28.32 MeV. That means  energy has to be 
transformed accordingly when using a cross section for reaction $e_{e2}^{-}+e_{e1}^{-}$  as shown
in Fig. 2.   
\subsection{Intensity of the electron beams and luminosity } 
\label{sec:10}
The knowledge of the number of electrons/s in the beams e1 and e2  provides the last ingredients for formula (1)
in order to calculate the luminosity.
Conservative numbers for electron currents with $I_{e1}$=$I_{e2}$ =10 mA  will do it at this point of 
discussion. In section 5
values already reached  will be addressed.  
With all numbers in place the luminosity reads 
\begin{equation}
L=\frac{6 \cdot 10^{16}\cdot 6 \cdot 10^{16}}{4\pi \cdot \ 0.3 \cdot 10^{-2} \cdot 0.3 \cdot 10^{-2} }%
=3.2 \cdot 10^{37}cm^{-2}\cdot s^{-1}
\end{equation}
That value for the luminosity L for one interaction zone has to be 
multiplied with the product of the periods of the beams e1 and e2 in the solenoid.
With that product 78$\cdot 5$ the luminosity of formula (5) increases to
\begin{equation}
L^{foci}=1.24 \cdot 10^{40}cm^{-2}\cdot s^{-1}
\end{equation}
The last step towards a calculation of positron rates
needs a knowledge of the cross section. 
\section{Cross sections and reaction rates }
\label{sec:11}
\subsection{Positron production }
\label{sec:12}
A cross section for the reaction $e^{-}
+e^{-}\rightarrow e^{-}+ e^{+}+ e^{-}+ e^{-}$ has been
reported in reference  \cite{gryaz}. The formula
\begin{equation}
 \sigma_{tot} =5.22 \cdot ln^{3}(\frac{2.3+E_{e^-}}{3.52}) \mu b
\end{equation}
has been extracted as a fit to the results of two previous publications  \cite{baier} and \cite{landau}.
The energy of beam e2, $E_{e2}$=10.01 MeV, corresponds to $E_{e2}^{e1(rest)}$=28.32 MeV in the laboratory system.
The total cross section $ \sigma_{tot}$ for that energy yields, using formula 7, $ \sigma_{tot}^{e-e}=52.8 \mu b$.   
Thus, the rate  of the number of positrons $N_{positron}$ results from the product luminosity times cross section: 
\begin{equation}
N_{positron}=L^{foci} \cdot \sigma_{tot}^{e-e}=6.55 \cdot 10^{11} s^{-1}
\end{equation}  
A discussion of uncertainty of the value of $ \sigma_{tot}^{e-e}$ could not be found. However, the value
has the right order of magnitude as estimates along the lines of virtual photon theory suggest.
\subsection{Other scattering processes}
\label{sec:13}
In addition to pair production two other reactions take place: Elastic electron scattering 
and bremsstrahlung. Elastic scattering is by far the largest contribution of all three processes. Fig. 8 
shows the momenta of the scattered electrons.
The Moeller cross section, however, has two poles from the t and the u channel
as well contributing to the cross section. The differential cross section \cite{PDG}, \cite{halzen}
\begin{equation}
\frac{d \sigma}{dt}=\frac{\pi \cdot  \alpha ^{2} \cdot \hbar ^{2}}{2\cdot s\cdot pe2cm^{2}}\cdot [\frac{s^{2}+u^{2}}{t^{2}}+\frac{2\cdot s^{2}}{t\cdot u} 
+\frac{s^{2}+t^{2}}{u^{2}}] 
\end{equation}
show forward, interference and backward pieces, with $\alpha$ as fine structure constant, $\hbar$ Planck constant divided by $2\pi$, s, t, u  Mandelstam variables and $pe2cm$ c.m. momentum of beam e2.
The total cross section, that means the integration of formula (9), diverges.  How to go around the singularities in principle, see \cite{feynman}. An integration over the range of positron angles $ 180^{0} \geq \vartheta _{positron} \geq 20^{0}$ yields still
a value of 0.5 b . 
The size of elastic cross section drops proportional to s, the positron production rate increases with the incoming electron energy.
Thus elastic scattering will be one of several boundary conditions by selection of an optimal set-up. The bremsstrahl
contribution plays a role, too, not so much as an additional electron background, but as a real photon beam into direction of the beam lines of the incoming beams.
\begin{figure}[h!]
\centering
\includegraphics[scale=0.55]{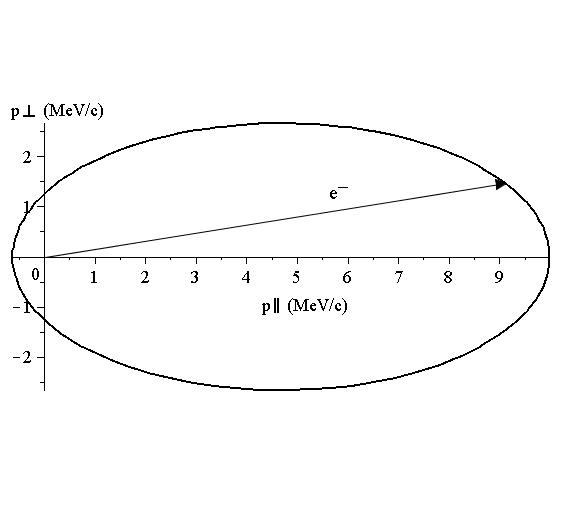}
\caption{The momentum space of the elastically scattered electrons via the reaction $e^{-}+e^{-} \rightarrow e^{-}+e^{-}$.}
\label{fig:8}
\end{figure}
\section{Polarized positrons}
\label{sec:14}
 An additional advantage using  pair production via photo- or electro induced reactions, besides reaching
 high rates, resides in the possibility to create polarized positrons. Already very early \cite{zel} it has been
 realized that longitudinally polarized electrons emit bremsstrahlung with circular polarization.
 Olson and Maximon \cite{maximon} extended that finding by 
showing that pair production induced with circularly polarized photons produce polarized positrons.
Finally, based on the general method described e.g. in \cite{jack}, first devised by Weizsaecker-Williams \cite{weiz} and \cite{wil}, tractable
expressions could be found.  Formulae for virtual photon spectra, see e.g. \cite{dalitz,tiator}, have been developed 
in order to understand and describe quantitatively electro induced reactions.The region close to the endpoint, that means highest energy point of the virtual photon spectrum, has been identified as energy region to carry out reliable calculations  in order to extract photo production cross sections out of electro production data, see e.g. \cite{schmidt}. 
Thus, along those lines it can be understood and calculated that circularly polarized electrons produce polarized positrons. 
Given the fact that electron beams with polarization $P_{e}\geq 80 \% $ are routinely used at several laboratories that method
can be applied without further difficulties.
The degree of polarization transfer depends
on the (virtual) photon energy transfer as shown in Fig. 9.
The helicity transfer for (virtual) photons close to the end point of the spectrum comes close to 100$\%$.
\begin{figure}[h!]
\centering
\includegraphics[scale=0.6]{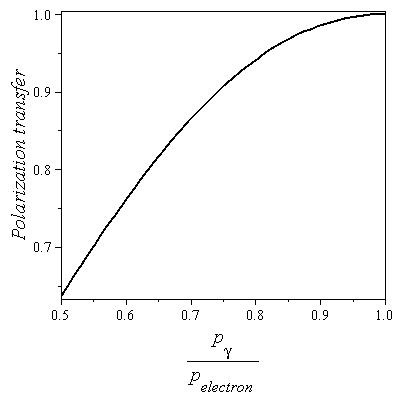}
\caption{Transfer of polarization from electron to virtual $\gamma $ to positron as a function of the relative $\gamma $ energy within a virtual photon spectrum using the formula from Ref. \cite{maximon}. $p_{\gamma}$ stands for the momentum of the virtual
photon and $p_{electron}$ for the momentum of the incoming electron.}
\label{fig:9}
\end{figure}
Thus, positrons with high momenta, see Fig. 7, have a high polarization close to the polarization of the incoming
electrons.
\newline 
Calculations within the framework of virtual photon theory for the given case show that 
the integrated intensity of the
virtual photon spectrum from threshold to the highest energy yield a polarization of positrons of 52 $\%$, 
integration of the upper half of the spectrum result in an intensity of 44$\%$ and a polarization of 76$\%$ and, 
finally, integration of the 
upper third of the virtual spectrum yield an intensity of 27$\%$ and a polarization of 85$\%$. Those values for the
polarization have to be multiplied with the degree of polarization of the electrons in order to get the final 
polarization
of the positrons.
The low mass of the target might lead to corrections \cite{tiator}. However, 
the corrections, especially close to the end point region, are probably not so large in order to change the picture 
described above. The closeness of the  cross sections of real photons on the proton and the electron shown in Fig. 1 
suggest, as already mentioned above, that the target mass plays not a dominant role in that reaction.  
\section{Developments towards higher electron currents}
\label{sec:15}
One of the useful properties of the method resides in a lesser dependence on the emittances of the beams in
order to achieve high luminosities compared to intersecting beams e.g. using storage rings. All particles entering the magnetic field parallel to the symmetry axis touch the axis. However, particles entering at a larger distance R to the symmetry axis do have a smaller $\beta _{z}$ and , thus, a larger trajectory s, see Appendix. Thus, part of the transverse emittance gets transferred towards the longitudinal direction. That property helps to use higher beam currents with poor transverse emittance in order to increase luminosity.
Remarkable progress has been achieved recently by preparing high electron beam current \cite{gulli},\cite{dun}.  Electron  
beam currents $I_{e}$ of up to $I_{e}$=100 mA have been achieved with a remarkable low emittance by using a d.c. gun 
with a voltage of 385 keV. The situation concerning polarized electrons looks
promising, too. Developments reaching currents in the mA region are under way \cite{grames}. An ambitious project 
tries to reach
a polarized electron beam current of $I_{e}$=50 mA by using a Gatlin gun \cite{gass},\cite{rah}. For such a gun the expected
normalized emittance $\epsilon _{n}$ of the beam is of the order of $\epsilon _{n}$=20 mmmrad.
\section{Conclusions}
\label{sec:16}   
Electro production of electron-positron pairs on electrons has the potential to provide
an efficient method to prepare positron beams. The method offers flexibility due to many parameters
to prepare a positron beam for a plethora of applications.  
Positrons can be produced without disturbances due to nuclear material causing multiple
scattering and producing - depending on energy - as a byproduct a large neutron background.
A high degree of polarization of positrons can be reached by using a
polarized electron beam and by selecting the upper part of the
momentum distribution of positrons.  
Calculations using the envelope equation, see e.g. \cite{rei}, will allow to judge the limits of the hard edge
approximation for both electron beams.

\section{Appendix} 

\label{sec:17}

\subsection{Beam e1 and e2}

\label{sec:18}
\vspace{1pt}

An electron gun delivers a beam e1 with a kinetic energy of $%
E_{kin}=300$ keV, an emittance of $\varepsilon _{n}=5$ mmmrad and a diameter 2R=2 mm. The total
energy $E1=0.81$ MeV, $ \gamma =\allowbreak 1.\,\allowbreak 59, $ the momentum $%
p1_{in}=\allowbreak 0.63$ MeV/c, $ \beta _{in}=\allowbreak 0.78$ and $p_{\Phi }=%
\frac{e\cdot r\cdot B_{z}}{2}=-\frac{300\cdot r\cdot 1}{2}=\allowbreak
-150.0\cdot r$ determine the trajectories of $\ $the electrons. $B_{z}=1.0$ $%
T$ has been chosen for \ the case considered here. For particles on the
periphery (R=1 mm) of the incoming beam the Lenz momentum $p_{\Phi }=\frac{%
300\cdot 10^{-3}\cdot 1}{2}=\allowbreak 0.15$ MeV/c and the radius of the
helix $r_{h}=\frac{p_{\Phi }}{300\cdot 1}=\frac{0.15}{300\cdot 1}%
=\allowbreak 0.000\,5$ m$=\frac{R}{2}.$ The cyclotron frequency is given by $%
^{{}}\omega =\frac{e\cdot B_{z}}{\gamma \cdot m}=\allowbreak 1.\,\allowbreak
11\times 10^{11}$ rad/s. The helix does have a different velocity in the
z-direction for each r due to $p_{\Phi }(r).$ Hence, $\beta _{z}$ changes,
e.g. for r=R, to $\beta _{z}=
0.755.$\vspace{1pt}The length of a period $l_{p}$ is given by $l_{p}=\tau \cdot
\beta _{z}\cdot c=\allowbreak 0.0128$ m, with $\tau =\allowbreak
5.\,\allowbreak 66\times 10^{-11}$ s given by the cyclotron frequency. The length of the helix $s(t,k)=2\pi
\cdot r\cdot \sqrt{1+k^{2}}\cdot t$ is determined by $l_{p}$ via slope $k=%
\frac{l_{p}}{2\pi \cdot r}$ and number of periods $t$ with $t=\frac{1}{l_{p}}%
=\frac{1}{0.0128}=\allowbreak 78.12$ for a effective field length of 1 m. The
slope for the peripheral electrons of the beam with $r=\frac{R}{2}=\frac{%
1\cdot 10^{-3}}{2}=\allowbreak 0.000\,5 m$ yields $k=\frac{0.0128}{2\pi \cdot
5\cdot 10^{-4}}=\allowbreak 4.\,\allowbreak 08.$, translating into an angle $%
\alpha $ of the helix of $\alpha =\arctan (\frac{2\pi \cdot 5\cdot 10^{-4}}{%
0.0128})=\allowbreak 13.\,\allowbreak 76%
{{}^\circ}%
.$ Putting the different pieces together the length of the helix amounts to $%
s(t,k)=\allowbreak 1.\,\allowbreak 0316\,$ m. With the chosen $%
\varepsilon _{n}$, the diameter 2R and the momentum of the incoming beam $%
p1_{in}$ the transverse momentum $p_{\perp }$ can be calculated to $p_{\perp
}=\frac{\varepsilon _{n}\cdot p1_{in}}{\gamma }=\frac{5\cdot 10^{-3}\cdot
0.63}{\allowbreak 1.\,\allowbreak 59}=\allowbreak 1.\,\allowbreak
981\,132\,1\times 10^{-3}$ MeV/c. That momentum has to be added to $p_{\Phi
}=\allowbreak 0.15$ MeV/c and leads to an extension of R=1 mm to $r=\frac{%
(0.15+0.001\,98)\cdot 2}{300}=\allowbreak 0.001\,013\,$ $m$. The trajectory
of electrons of the periphery of the beam contains the magnetic axis with a
distance of $\Delta r=13 \mu  $ m.

The beam e2 has an momentum $p2_{in}$=10 MeV/c, a beam diameter of 2�R = 2 mm, 
and normalized emittance $ \epsilon_{n}$=5 mmmrad. With those values the parameters 
for the trajectories can be calculated with the same formulae as used for beam 1. 
The results are shown together with the values of beam e1 in tables 1-4.

\subsection{Tables }

\label{sec:19}

\subsubsection{Common values for both beams}

\label{sec:20}

\begin{tabular}{||c||c||c||c||}\hline\hline
& $B_{z}$ & $p_{\Phi }$ & $r_{h}$ \\ \hline\hline
& T& MeV/c & mm \\ \hline\hline
$e1\&e2$ & 1 & 0.15 & 0.5 \\ \hline\hline
\end{tabular}

\subsubsection{Kinematic values}

\label{sec:21}

\begin{tabular}{||c||c||c||c||c||c||}
\hline\hline
& E & p & $\gamma $ & $\beta $ & $\omega $ \\ \hline\hline
& MeV & MeV/c &  &  & rad/s \\ \hline\hline
e1 & $0.81$ & $0.63$ & $\allowbreak 1.\,\allowbreak 59$ & $0.78$ & $%
1.\,\allowbreak 11\times 10^{11}$ \\ \hline\hline
e2 & $10.\,\allowbreak 01$ & $10.0$ & $19.\,\allowbreak 6$ & $0.999$ & $%
8.\,\allowbreak 975\times 10^{9}$ \\ \hline\hline
\end{tabular}

\subsubsection{Derived values 1}

\label{sec:22}

\begin{tabular}{||c||c||c||c||c||}
\hline\hline
& $\beta _{z}$ & $\tau $ & $l_{p}$ & $t$ \\ \hline\hline
&  & s & m &  \\ \hline\hline
e1 & $\allowbreak 0.755$ & $\allowbreak 5.\,\allowbreak 66\times 10^{-11}$ & 
$\allowbreak 0.0128\allowbreak $ & $78$ \\ \hline\hline
e2 & $\allowbreak 0.998$ & $7.\,\allowbreak 0\times 10^{-10}$ & $0.21$ & $%
4.\,\allowbreak 77$ \\ \hline\hline
\end{tabular}

\subsubsection{Derived values 2}

\label{sec:23}
 
\begin{tabular}{||c||c||c||c||c||c||}
\hline\hline
& $k$ & $s$ & $\alpha $ & $p_{\perp }$ & $\Delta r$ \\ \hline\hline
&  & m & deg & MeV/c & $\mu m$ \\ \hline\hline
e1 & $4.\,\allowbreak 08$ & $1.\,\allowbreak 032$ & $13.8$ & $1\,.98\cdot
10^{-3}$ & $13$ \\ \hline\hline
e2 & $\allowbreak 66.\,\allowbreak 75$ & $1.\,\allowbreak 000\,1$ & $%
0.\,\allowbreak 86$ & $2.55\cdot 10^{-3}$ & $17$ \\ \hline\hline
\end{tabular}

\end{document}